\documentclass[useAMS,usenatbib]{mn2e}
\def\ro{{\it ROSAT\/}}

\def\xmm{{\it XMM-Newton\/}}
\def\cha{{\it Chandra\/}}

\def\apj{{\it ApJ\/}}
\def\mnras{{\it MNRAS\/}}
\def\apjs{{\it ApJS\/}}
\def\aj{{\it AJ\/}}
\def\aa{{\it A\&A\/}}
\def\aas{{\it A\&AS\/}}
\def\g107{G107.5--1.5}

\usepackage{natbib}

\usepackage{color}

\usepackage{epsfig}
\usepackage{url}
\usepackage{amssymb}
\title{\xmm\ and Canadian Galactic Plane Survey Observations of the Supernova Remnant \g107}
\author[M. S. Jackson, S. Safi-Harb, \& R. Kothes]{M. S. Jackson$^{1~2~3}$\thanks{E-mail:mirandaj@kth.se},
S. Safi-Harb$^{1~4}$, \& R. Kothes$^{5}$\\
$^{1}$Department of Physics and Astronomy, University of Manitoba, Winnipeg, MB, R3T 2N2, Canada\\
$^{2}$Department of Physics, KTH Royal Institute of Technology, Stockholm, Sweden\\
$^{3}$The Oskar Klein Centre for Cosmoparticle Physics, AlbaNova University Centre, Stockholm, Sweden\\
$^{4}$Canada Research Chair\\
$^{5}$Dominion Radio Astrophysical Observatory, National Research Council Herzberg, P.O. Box 248, Penticton, BC, V2A 6J9, Canada}

\begin{document}
\date{Accepted . Received ; in original form 2013 July 19}


\maketitle

\label{firstpage}

\begin{abstract}
We present an \xmm\ observation of the highly polarized 
low-surface brightness supernova remnant \g107, discovered with the Canadian Galactic
Plane Survey (CGPS). We do not detect diffuse X-ray emission from the SNR and set an upper limit on the surface brightness of $\sim2 \times 10^{30}$~erg~arcmin$^{-2}$~s$^{-1}$, at an assumed distance of 1.1~kpc. We found eight bright point sources in the field, including the 
\ro\ source 1RXS J225203.8+574249 near the centre of the radio shell. Spectroscopic analysis of some of the
embedded point sources, including the \ro\ source, has been performed, 
and all eight sources are most likely ruled out as the associated neutron star, primarily due to counterpart bright stars in optical and infrared bands.
Timing analysis of the bright point sources yielded no significant evidence for pulsations, but, due to the timing resolution, only a small part of the frequency space could be searched. An additional ten fainter point sources were identified in the vicinity of the SNR. Further X-ray observation of these and the region in the vicinity of the radio shell may be warranted.

\end{abstract}


\section{Introduction}
Low-surface brightness supernova remnants (SNRs) are important in shaping our current
understanding of pulsar and supernova evolution, since
 the population of SNRs in our Galaxy is likely
to be dominated by faint objects.
In fact, around 85\% of Galactic SNRs are expected to be of core-collapse origin \citep{tsu95}. 
These are explosions of massive stars, most of which create stellar
wind bubbles around them, and thus form low-surface brightness SNRs.
Massive star explosions should also lead to the formation of a
compact object that could be revealed in the radio and/or X-rays.
High-resolution \cha\ and \xmm\ observations have been particularly crucial in shedding light on
the collapsed cores of the explosions and their relativistic outflows \citep[see][]{kar08,vink12}. 

In particular, high-resolution and sensitive X-ray observations have contributed to the discovery of a diversity
of neutron stars in SNRs. This diversity includes rotation-powered pulsars such as the Crab and Vela pulsars, 
magnetically powered pulsars (dubbed as `magnetars' and including the Anomalous X-ray Pulsars and Soft Gamma-ray Repeaters), and other X-ray emitting neutron stars whose nature is still being debated \citep[see e.g.][]{mer12}. These include the Central Compact Objects (CCOs) in SNRs and are typified by the CasA compact object \citep[e.g.][]{got07}. Finding new compact objects in SNRs will therefore help in shedding light on this diversity and on the connection between the compact objects and the progenitor stars forming their hosting SNR.

The Canadian Galactic Plane Survey (CGPS) \citep{tay03}
is a high-resolution radio survey that incorporates single-antenna
data to retain sensitivity to the largest
structures (such as faint filamentary non-thermal radio emission from SNRs).
In 2001, the shell-type supernova remnant (SNR), \g107, was
discovered in CGPS data \citep{kot03}. 
Including the one described in this paper, eight SNR candidates have so far been discovered in the CGPS 
\citep{kot01,kot05,tia07,fos13}.
In 2008, we started a program to search for these SNRs
and their associated compact objects in \cha\ and \xmm\ data \citep[see e.g.][]{jac08}.

\begin{figure}
\includegraphics[height=3in]{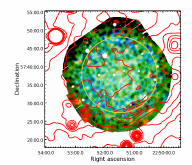} 

\caption{False colour \xmm\ mosaic image of SNR \g107. Sources have been excised and the image has been smoothed with a threshold of 50 counts.
The orange component corresponds to 0.5--2.5~keV and the cyan component corresponds to 2.5--10~keV. The energy bands were chosen so that
each of the component images has an approximately equal number of counts. The images have had particle and soft proton backgrounds subtracted and are exposure-corrected. The radio contours are overlaid in
 red. The extraction region used for the spectral analysis in \S\ref{diffspec} is shown in yellow. Coordinates are J2000.
\label{xrayradioimage}}
\end{figure}

The SNR G107.5--1.5 is located
in a complex region of the Galactic plane
$\sim$4 degrees west of Cas~A \citep{kot03}. At 1420 MHz, the source appears as a thin filament on top of diffuse shell-like
emission (see contours in Figure~\ref{xrayradioimage}).
Combining the 408 MHz and the 1420 MHz emission, the spectral
index was determined to be $\alpha$ $\sim$ 0.6 ($S_{\nu}$ $\propto$
$\nu^{-\alpha}$) -- typical of shell-type SNRs. 
Its identification as an SNR is based mainly on its high
polarized emission coincident with the filament and spectral index.
Its polarized intensity is 50\% at 1420 MHz, and
its peak fractional polarization  of 71\% is close to the theoretical maximum
(for $\alpha$=0.6).
This was therefore identified as the most highly polarized SNR known.
The distance to this SNR from various
distance determination methods \citep[as described in][]{kot03} is $1.1\pm0.4$~kpc. The morphology of
the SNR and the structure of the ambient neutral hydrogen in the radio
data suggests that this SNR is $\sim$3--6~kyr-old, and believed to be 
 in a late stage of evolution, given that it is expanding in a high ambient density medium or originates from a supernova with low explosion energy \citep{kot03}.

\begin{figure*}

\includegraphics[height=6in]{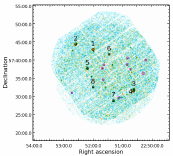}

\caption{False colour \xmm\ mosaic image of SNR \g107. Energy bands, background subtraction, and exposure-correction are the same as in Figure~\ref{xrayradioimage}.
The image has been smoothed with a threshold of 20 counts. Detected point sources are circled, and those from which spectra could be extracted are numbered \textbf{1--8 and identified with black circles.}  
\label{xrayradioimageps}}
\end{figure*}

At the projected centre of the SNR shell, there is a \ro\
unidentified source, 1RXS J225203.8+574249,
whose location is quite compelling:
not only is it close to the projected centre of the SNR,
but if it is a neutron star and the explosion happened at its current
position, then it would explain the peculiar appearance of the SNR as a shell fragment. In
all directions, the wind of the progenitor and the shock wave
of the explosion would have been blocked by dense neutral hydrogen except for the
gap in the direction of the radio shell. This, together with the SNR's
evolutionary state and location in a complex region of the Galaxy,
leads us to believe that the supernova explosion \textbf{was of core-collapse type},
and the \ro\ source, \textbf{located close to source 1}
in Figure~\ref{xrayradioimageps}, 
is possibly the collapsed core.

The small number of
counts in the \ro\ data did not allow, however, for spectroscopy or detailed
imaging of the overall region or point sources to be done, so an \xmm\
observation, described in \S\ref{obs}, has been carried out, primarily in order to
make an X-ray detection of the SNR, and to study the \ro\ source.

Because the integration time was unfortunately significantly reduced by proton
flaring (see \S\ref{obs}),
distinguishing diffuse X-ray emission from the SNR on the images was not possible, but we estimate the strength of its X-ray emission through spectral analysis.
In this work, the main focus is to examine the discrete 
sources in the field. We were able to resolve a few dozen sources, \textbf{18 of which have more than 25 counts according to the source detection}, and extract spectra
for eight of them, including two very close to the location of the \ro\ source, which we preliminarily consider to be the likely neutron star associated with \g107.

\section{Observation}\label{obs}
The SNR \g107 was observed with \xmm\ in full window mode on 2007
January 23 for 34.8 ks (Obs ID: 0400600101) The pn \citep{str01} and
MOS \citep{tur01} data were reduced with \textbf{Scientific Analysis System (SAS)}\footnote{\url{http://xmm.esa.int/sas/}} version 13.5.1.

Images and spectra were extracted using the \textbf{\xmm\ Extended Source Analysis Software (ESAS)} package \citep{esas}, which calls various SAS routines to filter out proton flaring, identify and remove point sources (for the diffuse analysis), and estimate backgrounds for the diffuse spectra by using the non-illuminated corners of the detectors. Because of the high amount of proton flaring, the integration time was reduced to 11.7~ks for pn, 15.7~ks for MOS1, and
15~ks for MOS2.


\section{Imaging}\label{imaging}
Figures~\ref{xrayradioimage} and \ref{xrayradioimageps} show the \xmm\ 
X-ray mosaic images in the 0.5--10 keV range, with CGPS contours overlaid. These are combined, exposure-corrected images from 
pn, MOS1, and MOS2 data from which particle and soft proton backgrounds have been subtracted. 
Many discrete point-like sources and
a few extended sources, most of which lie
within the radio shell, can be seen on Figure~\ref{xrayradioimageps}. 

Two methods were used to remove the contribution of the detected point 
sources (see \S\ref{detection} for a description of the source detection) to facilitate a close examination of the diffuse X-ray emission.
For the first, the reconstructed image of the detected sources produced 
by the source detection program was subtracted from the original image. The second method is to exclude regions at the locations of the detected sources, which is done using the ESAS routine {\it cheese}.

Images generated using the above-mentioned methods show no discernible diffuse X-ray emission (see Figure~\ref{xrayradioimage}).
This is possibly
because the exposure time was significantly shortened (by more than 50\%) due to the proton flares during
the observation. 
The objective of this paper is therefore to address the
nature of the point sources, with specific focus on source 1, which is
very near the location of the \ro\ source. However, an attempt has been made to identify areas of SNR X-ray emission through spectral analysis. This effort is described in \S~\ref{diffspec}.

\subsection{Source Detection}\label{detection}


To identify the X-ray sources in the \xmm\ observations (both point sources and extended
sources) in the SNR G107.5$-$1.5
field, {\it ewavelet}, a wavelet detection algorithm which is part of
the SAS package, was used. For each source, the output of the
routine gives the position on the image, source
counts, and source extent (FWHM), as well as errors in those quantities. The program
also produces a reconstructed image of the detected sources, which can be subtracted from the 
original image to reveal any diffuse emission, as described in the previous section. 
Point sources are identified by choosing sources at all wavelet scales with a significance of greater than 3$\sigma$ and a less than 5 pixel extent (FWHM), which corresponds to about 17 arcseconds, on the combined 0.5--2.5~keV and 2.5--10~keV images. This conservative requirement is based on the fact that off-axis sources have a larger PSF than those toward the centre of the images. This effect is more pronounced at higher energies, and the 5 pixel extent limit corresponds to the approximate PSF of a source at the highest energy at the edge of the field of view. The 18 sources with more than 25 counts from the source detection are shown in Figure~\ref{xrayradioimageps} as \textbf{magenta} or black circles. Only source 3 was significantly detected on both the soft and hard images. The other sources were detected only in the soft-band image, with no detection or a detection of fewer than 25 counts on the hard image. Source detection performed on the individual pn, MOS1, and MOS2 images in those energy ranges yielded similar source lists. The 2XMMi catalogue \citep{wat09} contains the 18 sources described above, as well as 43 fainter point sources within the region covered by the observation.

Eight of the eighteen point sources indicated on Figure~\ref{xrayradioimageps} are sufficiently bright to have spectra of more than 100 counts
extracted, and these are indicated as the numbered sources in Figure~\ref{xrayradioimageps}. Table~\ref{pscat} lists these sources with their \textbf{coordinates and positional errors.}
Source 1, which is near the centre of the radio shell,  has coordinates 
$\alpha$(J2000) = 22h~52m~00s.629, $\delta$ (J2000) = 57$^\circ$~42$'$~59$''$.00, with a positional error
of 0$''$.5, and is coincident with the
unidentified \ro\ source 1RXS J225203.8+574249 and is furthermore identified as 2XMMi J225200.3+574259 (see Table \ref{pscat}).

\begin{table*}
\centering
\begin{minipage}{140mm}
\caption{Catalogue of X-ray sources detected inside SNR G107.5--1.5, with corresponding 2XMMi name \citep{wat09}. The positional error is $1\sigma$ and corresponds to that found in the 2XMMi catalogue. The coordinates quoted for this paper give the locations of the centroids of the extraction circles for the point sources. \label{pscat}}
\begin{tabular}{lcccccccc}
\hline
\hline
\# & IAU Name &\multicolumn{2}{c}{This paper}&\multicolumn{2}{c}{2XMMi catalogue}& Pos. Err.&Spectral counts& Hardness ratio\protect{\footnote{The hardness ratio is calculated from the counts derived from the source detection on the soft and hard images and is calculated as (H-S)/(S+H).}} \\
&2XMMi& $\alpha_{\rm J2000.0}$ & $\delta_{\rm J2000.0}$ & $\alpha_{\rm J2000.0}$ & $\delta_{\rm J2000.0}$&(arcsec)&&\\
\hline
\hline
1&  J225200.3+574259& 22 52 00.63& 57 42 59.0&22 52 00.41& 57 42 59.0& 0.5&667&-1\\ 
2&  J225236.3+574422& 22 52 36.58& 57 44 21.7& 22 52 36.31& 57 44 22.2& 0.6 &714&-0.71\\ 
3&  J225036.9+573154& 22 50 37.05& 57 31 53.3&22 50 36.94& 57 31 54.5& 0.6&256&-0.28\\ 
4&  J225037.9+573127& 22 50 38.01& 57 31 27.4&22 50 37.97& 57 31 27.1& 0.6 &456&-1\\
5&  J225212.1+573744& 22 52 12.25& 57 37 43.8&22 52 12.19& 57 37 44.8& 0.6 &398&-0.21\\ 
6&  J225127.4+574136& 22 51 27.54& 57 41 36.5& 22 51 27.43& 57 41 36.2& 0.8 &192&-1\\
7&  J225118.7+572848& 22 51 19.02& 57 28 51.4& 22 51 18.79& 57 28 48.4& 0.9 &103&-0.09\\ 
8&  J225200.9+573230& 22 52 00.87& 57 32 30.7& 22 52 00.96& 57 32 30.8& 0.9 &106&-0.32\\ 
\hline
\end{tabular}
\end{minipage}
\end{table*}

Figure~\ref{xrayradioimageps} shows a false-colour image of
the sources. Orange corresponds to 0.5--2.5~keV and cyan corresponds to 2.5--10~keV.
\textbf{This image provides a rough indication of the spectral hardness of the sources.} 
Table~\ref{pscat} includes a column 
giving the total counts in the spectrum for each source.
Sources 1, 2, 4, and 6 appear softer than the others on the image. Sources 3 and 4, though very close to each other, were
detected as two distinct point sources, and \textbf{they clearly have different hardnesses.} 

The {\it Vizier} tool \citep{och00} was used to search for optical and infrared
counterparts to the X-ray sources. 
The
optical/X-ray flux ratio has been calculated for each counterpart candidate, which is
meaningful and accurate only if the counterpart candidate is the
true counterpart, and is the only (or at least the brightest) X-ray source within the extraction circle. As can be seen from Table~\ref{optcat}, sources 3, 7, and 8 have no close optical counterpart candidate.

\begin{table*}
\centering
\caption{List of optical (O) and infrared (I) counterpart candidates for each numbered X-ray source. Only those counterpart candidates within the X-ray error circle are listed. 
All counterpart candidate designations are from the USNO-B1.0 catalogue \citep{mon03} or from the 2MASS catalogue \citep{2mass}.   
\textbf{The B (the B1 magnitude is listed in this table) and J magnitudes correspond to those found in the USNO-B1.0 and 2MASS catalogues, respectively.}
}
\begin{minipage}{140mm}
\begin{tabular}{cllccc}
\hline
\hline
Source \#& Counterpart \#& USNO-B1.0/2MASS name& Offset (arcsec)&B/J magnitude&$\displaystyle\frac{\rm Optical}{\rm X-ray}$ flux ratio \protect{\footnote{The (logarithmic) optical to X-ray flux ratio is calculated assuming that only the single optical counterpart contributes to the X-ray flux. 
\textbf{The visible flux was calculated by using the fact that the B1 magntude of Sirius is -0.09 \citep{mon03} and the visible flux is} $1.1\times10^{-4}$~erg~cm$^{-2}$~s$^{-1}$ \citep{lie05,per97,och00}. \textbf{An uncertainty of $\sim1$ on these values can be assumed.}}}\\
\hline
\hline
1&O1&1477-0514019&3.6&9.4&4\\ 
&I1&22520056+5743004&1.5&6.1\\
\hline
2&O1
&1477-0514478&2.9&15.5&2\\
&O2&1477-0514481&2.1&14.9&2\\
&I1&22523638+5744221&1.6&12.1\\
\hline
4&O1&1475-0504764&0.63&11.8&3\\ 
&I1&22503796+5731271&0.4&9.9\\
\hline
5&O1&1476-0513879&0.68&18.0&2\\
&I1&22521215+5737435&0.8&12.9\\
\hline
6&O1&1476-0513126&0.74&15.7&4\\
&I1&22512749+5741359&0.7&12.1\\
\hline
8&I1&22520093+5732294&1.4&17.1\\
\hline
\hline
\end{tabular}
\end{minipage}
\label{optcat}
\end{table*}




The blue magnitude is converted into optical flux with the
relation 
$F_1=F_2 \cdot 10^{(m_2-m1)/2.5}$
and compared with the X-ray flux
(obtained from the spectral fitting, described below), 
and this ratio is one factor that can be checked to favour identification
as a neutron star. In the above
equation, $m_1$ and $m_2$ are the blue magnitudes of the object and a
standard star, respectively, and $F_1$ and $F_2$ are their fluxes. Typical neutron
stars have a flux ratio in optical to X-ray bands of $\sim 1-10$ \citep{lyn98},
whereas stars have a minimum flux ratio of $10^3$ \citep{tes10}. 
X-ray emitting O- or B-type stars such as $\eta$ Carinae \citep{cor00} or $\tau$
Scorpii \citep{mew03} have a ratio of $\sim 10^7$ \citep[this calculation is confirmed in][by means of the use of bolometric optical fluxes]{naz11}. 


\section{Spectroscopy}\label{spec}
\subsection{Diffuse emission}\label{diffspec}
To estimate the diffuse flux from the SNR, spectra were extracted
from a single circular region with a radius of 10 arcminutes centered at ${\rm RA}=22{\rm h}51{\rm m}30.4{\rm s}$, $\delta=57^\circ37^\prime46^{\prime\prime}$
in pn and MOS data. \textbf{Regions surrounding detected sources were excluded from the extraction. }In addition, a \textbf{\ro\ All-Sky Survey (RASS) spectrum} from a 1--2 degree annulus was retrieved from the HEASARC X-ray background tool v2.5\footnote{\url{http://heasarc.gsfc.nasa.gov/cgi-bin/Tools/xraybg/xraybg.pl}}, and all four spectra were used for the fits.

Following the ESAS cookbook\footnote{ftp://xmm.esac.esa.int/pub/xmm-esas/xmm-esas.pdf}, a model comprising instrumental Gaussian lines, three thermal components corresponding to the local hot bubble and hotter and cooler galactic halo, and a power law representing the background of unresolved X-ray sources, was preliminarily constructed using XSPEC\footnote{\url{http://heasarc.gsfc.nasa.gov/xanadu/xspec/}} version 12.8. The non-instrumental background components were considered to have fixed parameters across the entire region, and an additional broken power law component that was not folded through the instrumental response was used to fit the soft proton background. This final component was given a break energy of 3.0~keV, and the photon indices for the two MOS instruments were tied together. In the pn spectra, the soft proton background was found to be negligible and was removed from the fit. The thermal components were modelled with the Astrophysical Plasma Emission Code (APEC) model \citep{smi01}, modified (if necessary) by Wisconsin absorption \citep[{\it wabs} in XSPEC;][]{morr83}.

In the model described above, the thermal component corresponding to emission from the local hot bubble is unabsorbed and has a temperature of $kT\sim0.1$. Both components representing the halo emission are absorbed with an $N_{\rm H}$ value in the direction of the SNR found from both the HEASARC Column Density tool \footnote{\url{http://heasarc.gsfc.nasa.gov/cgi-bin/Tools/w3nh/w3nh.pl}} and the {\it Colden} tool on the \cha\ website\footnote{http://cxc.harvard.edu/toolkit/colden.jsp} to be $\sim 6 \times 10^{21}$~cm$^{-2}$ from Galactic maps \citep{lab,dl,sta}.

Because the line of sight is so close to the galactic plane, the component representing the hotter galactic halo emission includes thermal contribution from the galactic disk. This means that the absorption coefficient for the component varies over the emitting region, and it must be considered that the average absorption coefficient (that which will produce the best fit for the model) may not be as great as the value listed above because some of the emission arises from regions interior to the galaxy. 

Unfortunately, the temperatures of the Galactic (halo and disk) emission and the SNR emission are expected to be similar ($\sim 0.25-0.7$~keV for the Galactic emission and $\sim 0.5$~keV for the SNR emission). In addition, the hydrogen column density, $N_{\rm H}$, in the direction of
the SNR, has been determined to be in a range of 0.1 to 0.6 $\times
10^{22}$~cm$^{-2}$. The lower limit on $N_{\rm H}$ was determined from
radio data which included only neutral atomic hydrogen \citep{kot03}, and the upper
value was determined from measurements of the absorption along the full line of sight through the galaxy as above. The upper value is also consistent, within error, with the radio determination. Because the temperatures and absorption coefficients of the two components are potentially indistinguishable, because the SNR emission is not discernible on the images, and because the spectral fitting did not allow for two separate components, the Galactic emission and the SNR emission have been combined into a single model component with a hydrogen column density up to $6 \times 10^{21}$~cm$^{-2}$ , and the maximum flux of this component can be considered to be the upper limit on the SNR emission.

The final model therefore includes instrumental lines, broken power law components representing the soft proton contamination in the MOS1 and MOS2 spectra, a power law representing the cosmic X-ray background and unresolved X-ray point sources, an unabsorbed thermal component, and two absorbed thermal components \textbf{(``cool'' and ``warm'') }corresponding to emission from the Galactic halo and disk and from the SNR. The higher temperature (warm) thermal component is likely to include all emission from the SNR.

The model was fit from 0.1 to 2.0~keV for \ro, from 0.4 to 10~keV for MOS1 and MOS2, and from 0.5 to 10~keV for pn to the model described above. The coefficients describing the hydrogen column density were fixed at $6 \times 10^{21}$~cm$^{-2}$ to better constrain the fits. The fit was of reasonable quality, with $\chi^2_\nu =1.05$. \textbf{The relevant parameters from the fit for the cool and warm thermal components are given in Table~\ref{diffspectable}. Given the size of the extraction region and the unabsorbed flux of the warm thermal component, }this indicates an unabsorbed surface brightness upper limit for the SNR of $\sim2\times10^{30}$~erg~arcmin$^{-2}$~s$^{-1}$.


\begin{table}
\caption{Fit parameters pertaining to the cool and warm thermal components in the model described in \S\ref{diffspec} to which spectra extracted from the yellow circle in Figure~\ref{xrayradioimage} have been fitted. Errors are 2$\sigma$ uncertainties. \label{diffspectable}}
\centering
\begin{tabular}{lccc}
\hline
\hline
Component&$kT$ (keV)&\multicolumn{2}{c}{0.5--5.0~keV Flux}\\
&&\multicolumn{2}{c}{($10^{-12}$erg cm$^{-2}$s$^{-1}$)}\\
&&Absorbed&Unabsorbed\\
\hline
\hline
Cool&$0.15^{+0.02}_{-0.01}$&$(0.4 \pm 0.2)$&$(18 \pm 7)$\\
\hline
Warm&$0.58^{+0.04}_{-0.07}$&$(0.6 \pm 0.1)$&$(3.5 \pm 0.6)$\\
\hline
\hline
\end{tabular}
\label{psfitbb}
\end{table}

\begin{figure}
\scalebox{0.33}{\rotatebox{270}{\includegraphics*{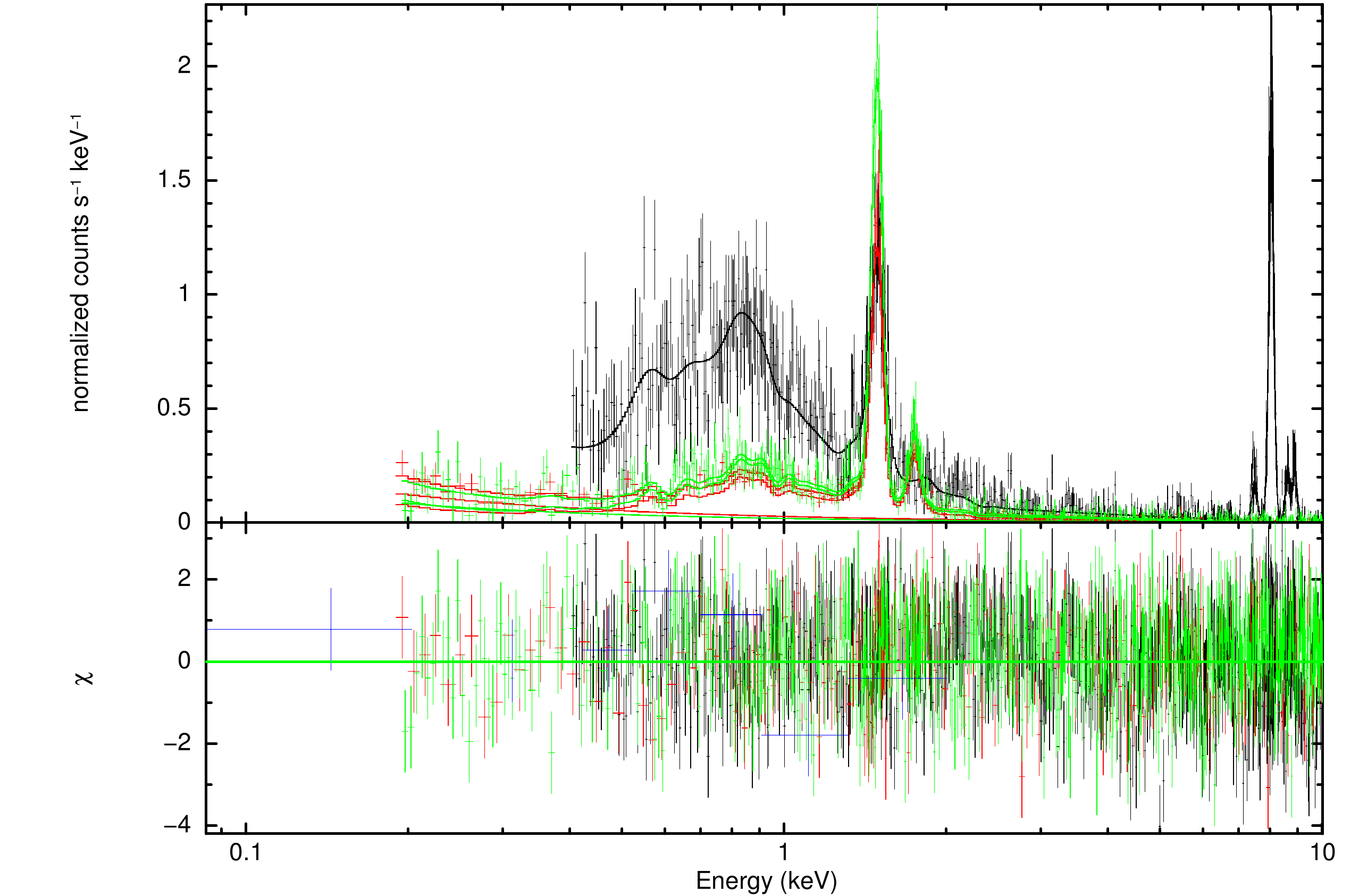}}}
\caption{Fit of spectra extracted from the yellow circle in Figure~\ref{xrayradioimage} to the model described in \S\ref{diffspec}. Black, red,
  green, and blue correspond to pn, MOS1, MOS2, and RASS spectra, respectively. The
  lower panel shows the deviation of the data from the model in terms of $\chi$.
\label{specdiff}}
\end{figure}

\subsection{Point sources}
\begin{figure}
\scalebox{0.33}{\rotatebox{270}{\includegraphics*{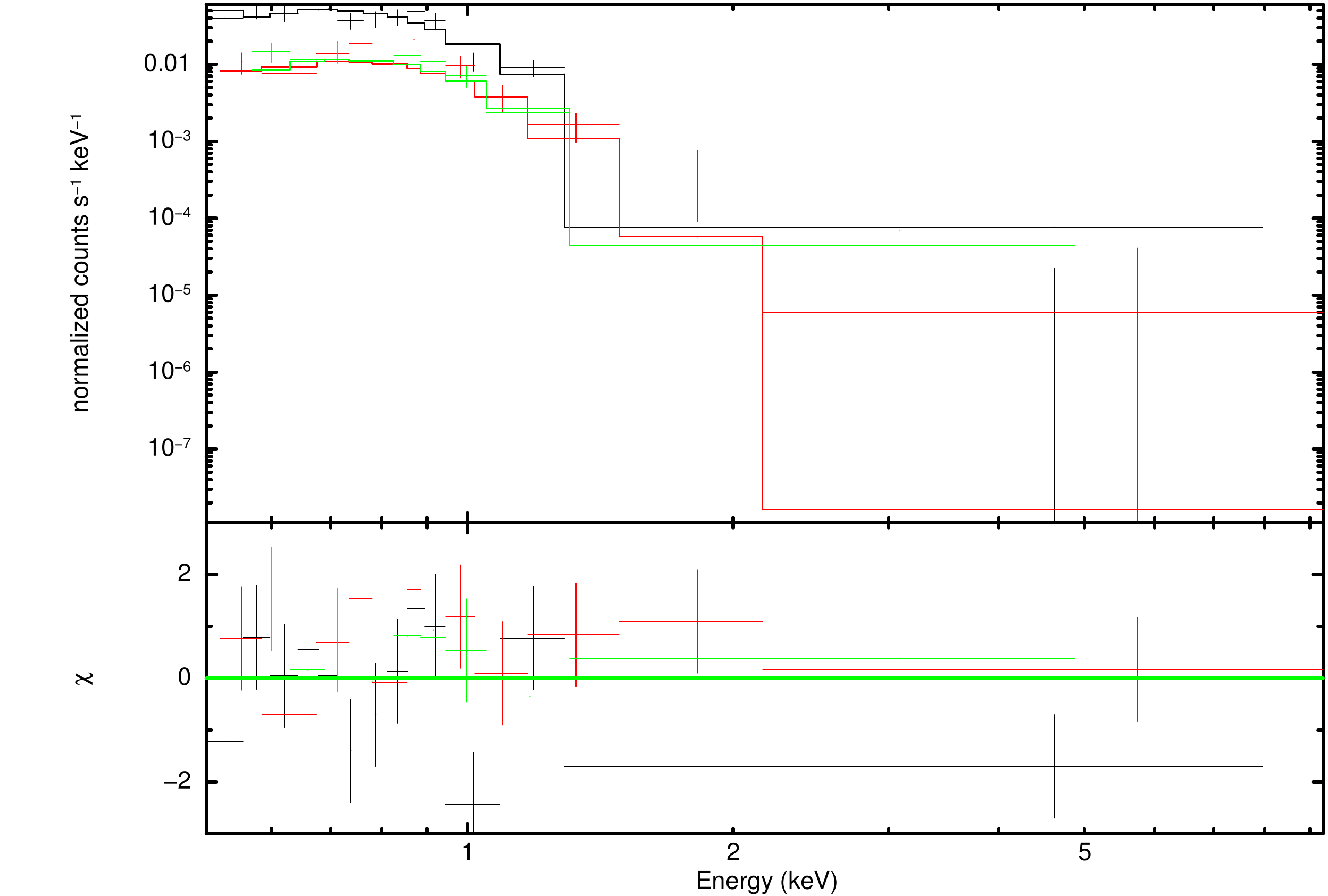}}}
\caption{Spectrum of source 1 fit to an absorbed blackbody. The fitted
  parameters are given in Table~\ref{psfitbb}. Black, red, and
  green correspond to pn, MOS1, and MOS2 spectra, respectively. The
  lower panel shows the deviation of the data from the model in terms of $\chi$.
\label{specsrc1}}
\end{figure}

For each numbered source in Figure~\ref{xrayradioimageps}, spectra have
been extracted from the pn and MOS data. The spectra have been grouped 
into bins containing at least 20 counts for pn and 10 counts for MOS. 
The background spectrum for
each source has been extracted from an annulus surrounding the source,
avoiding other nearby sources. The pn and MOS spectra for each
individual object have been simultaneously fitted to an absorbed
blackbody and/or an absorbed power law model. A likely identification for a Galactic X-ray point source with an infrared or optical  counterpart is a coronally active late-type star \citep{war11}, so it is necessary to investigate each source with that possibility in mind. Thus, each source has also been fitted to an APEC model modified by Wisconsin absorption, with the expectation that coronal emitters will have a temperature kT of 0.1--1 keV \citep{gud04}.
The results are
shown in Tables~\ref{psfitbb}, \ref{psfitapec}, and \ref{psfitpow} for those sources whose spectra are well fitted with these models, and Figure~\ref{specsrc1} shows the fit of the spectrum of source 1 to an absorbed blackbody. 

\begin{table*}
\caption{Fit of discrete source spectra to an
  absorbed blackbody. Parameters are listed only where a satisfactory fit could be obtained. Errors are 2$\sigma$ uncertainties.  }
\begin{minipage}{\textwidth}
\centering
\begin{tabular}{lccccc}
\hline
\hline
\# &$N_{\rm H}$ ($10^{22}$cm$^{-2}$)&$kT$ (keV)&0.5--5 keV Flux\protect\footnote{All flux values refer to unabsorbed flux.} ($10^{-12}$erg cm$^{-2}$s$^{-1}$)&$\chi^2_\nu$ ($\nu$)&Radius(km)\\
\hline
\hline
1&$0.3\pm0.2$&$0.10^{+0.03}_{-0.01}$& $0.3^{+1.5}_{-0.2}$&1.06 (31)\\ %
&(0.6)\protect\footnote{A number in parentheses indicates a frozen value for the fit.}&$0.082\pm0.004$&$2.1^{+0.7}_{-0.5}$&1.18
(32)&$20^{+5}_{-4}$\\ %
\hline
2&$0.3\pm0.2$&$0.10\pm0.03$&$0.3^{+2.0}_{-0.2}$&1.12 (28)\\ %
&(0.6)&$0.083\pm0.004$&$2.1^{+0.5}_{-0.4}$&1.24 (29)&$19^{+5}_{-4}$\\ %
\hline
3&$<0.14$&$1.0^{+0.2}_{-0.1}$&$0.09^{+0.03}_{-0.02}$&0.81 (20)\\ %
&(0.6)&$0.9\pm0.1$&$0.11\pm0.02$&1.33 (21)&$0.016^{+0.006}_{-0.005}$\\ 
\hline
4&$1.8^{+0.8}_{-0.7}$&$0.05\pm0.01$& (unconstrained)&0.76 (18)\\ %
&(0.6)&$0.09\pm0.01$&$1.1^{+0.7}_{-0.2}$&1.23 (19)&$11^{+5}_{-3}$\\ %
\hline
5&$<0.03$&$0.9\pm0.1$&$0.09\pm0.02$&1.77(20)&\\ %
\hline
6&$<0.24$&$0.26^{+0.06}_{-0.05}$& $0.016\pm0.003$&1.10 (6)\\ %
\hline
7&$<1.1$&$0.7\pm0.3$& $0.03^{+0.02}_{-0.01}$&0.37 (6)\\ %
&(0.6)&$0.6\pm0.2$& $0.03\pm0.01$&0.457 (7)&$0.019^{+0.017}_{-0.008}$\\ %
\hline
8&$<3.2$&$0.5^{+0.2}_{-0.3}$& $0.03^{+0.12}_{-0.01}$&0.41 (2)\\ %
&(0.6)&$0.4^{+0.2}_{-0.1}$& $0.04^{+0.02}_{-0.01}$&0.30 (3)&$0.03^{+0.05}_{-0.01}$\\ %
\hline
\hline
\end{tabular}
\end{minipage}
\label{psfitbb}
\end{table*}
\begin{table*}
\caption{Fit of discrete source spectra to an absorbed APEC. Parameters are listed only where a satisfactory fit could be obtained. Errors are 2$\sigma$ uncertainties.  }
\begin{minipage}{\textwidth}
\centering
\begin{tabular}{lcccc}
\hline
\hline \# &$N_{\rm H}$ ($10^{22}$cm$^{-2}$)&$kT$ (keV)&0.5--5 keV Flux ($10^{-12}$erg cm$^{-2}$s$^{-1}$)&$\chi_\nu^2$ ($\nu$)\\
\hline 
\hline
1&$0.61^{+0.07}_{-0.12}$&$0.15^{+0.03}_{-0.01}$&3$\pm2$&1.15 (31)\\ %
\hline
2&$0.62^{+0.07}_{-0.14}$&$0.15^{+0.04}_{-0.03}$&$3^{+3}_{-2}$&1.09 (28)\\ %
\hline
3&$0.4^{+0.4}_{-0.2}$&$>9$&$0.11^{+0.02}_{-0.03}$&0.91 (20)\\ %
\hline
4&$<0.09$&$0.69\pm0.08$&$0.036^{+0.008}_{-0.007}$&1.26 (18)\\ %
\hline
7&$<2.4$&$>0.9$&$0.04^{+0.12}_{-0.02}$&0.365 (6)\\ %
\hline
8&$<3.8$&$1.0^{+6.0}_{-0.7}$&$0.2^{+1.8}_{-0.1}$&0.29 (2)\\ %
\hline
\hline
\end{tabular}
\end{minipage}
\label{psfitapec}
\end{table*}

\begin{table*}
\caption{Fit of discrete source spectra to an
  absorbed power law. Parameters are listed only where a satisfactory fit could be obtained. Errors are 2$\sigma$ uncertainties.  }
\begin{minipage}{\textwidth}
\centering
\begin{tabular}{lcccccc}
\hline\hline 
\#&$N_{\rm H}$ ($10^{22}$cm$^{-2}$)&$\Gamma$&0.5--5 keV Flux\footnote{All flux values refer to unabsorbed flux.} ($10^{-12}$erg cm$^{-2}$s$^{-1}$)&$\chi_\nu^2$ ($\nu$)\\
\hline
\hline
3&$<0.64$&$1.0^{+0.5}_{-0.4}$&$0.10^{+0.10}_{-0.03}$&0.87 (21)\\%
&(0.6)\protect\footnote{A number in parentheses indicates a frozen value for the fit.}&$1.4\pm0.3$&$0.13\pm0.03$&0.94 (21)&\\%
\hline
5&$<0.1$&$1.0^\pm0.2$&$0.09^{+0.03}_{-0.01}$&0.97 (20)\\ %
\hline
6&$<0.71$&$3^{+4}_{-1}$&$0.03^{+0.1}_{-0.01}$&1.06 (6)\\ %
&(0.6)&$5.9\pm0.9$&$0.27^{+0.05}_{-0.04}$&1.16 (7)&\\%
\hline
7&$<2.5$&$2^{+3}_{-1}$&$0.06^{+0.9}_{-0.05}$&0.29 (6)\\%
&$(0.6)$&$2.0^{+0.7}_{-0.6}$&$0.05^{+0.04}_{-0.02}$&0.27 (7)\\%
\hline
8&$<5.4$&$>2.9$&$0.6^{+0.1}_{-0.4}$&0.46 (2)\\%
&(0.6)&$3\pm1$&$0.07^{+0.05}_{-0.03}$&0.48 (3)&\\%
\hline
\hline
\end{tabular}
\end{minipage}
\label{psfitpow}
\end{table*}


As was done with the diffuse spectrum, as described in \S~\ref{diffspec}, the $N_{\rm H}$ was estimated to be  0.6 $\times
10^{22}$ cm$^{-2}$ for sources associated with the SNR, and for
some of the blackbody and power law fits,  the value of $N_{\rm H}$ was
frozen at this value. This was 
done in
order to better constrain the 
fits, and to determine the spectral parameters for each source,
assuming an association with the SNR exists. Some spectra were not fitted well with the absorption parameter frozen in this way, and those are not shown in the tables.

When the $N_{\rm H}$ value was set to
0.6$\times 10^{22}$ cm$^{-2}$, the value of $\chi^2_\nu$
was greater than 2 for sources 5 and 6 for the absorbed blackbody fit, and source 5 for the absorbed power law fit, which indicates an unsatisfactory fit,
but it is a possible value of $N_{\rm H}$ for the rest of the
sources. The error bars on the parameters are unfortunately very large for many of the
sources, leading to large uncertainties in the flux
values. It is unclear for many of the sources which of the models, blackbody, APEC, or power law, is the best fit. 

For the objects fitted with an absorbed blackbody
model, the emitting radii can be determined, given the approximate
distance of 1.1~kpc. The radii are given in 
the rightmost column of Table~\ref{psfitbb} and those for sources 1, 2, and 4 are close to typical neutron star radii.

For many of the sources, the fit to an absorbed APEC is of comparable quality with the fit to an absorbed power law or blackbody, indicating that neither the neutron star nor the coronal emission explanation can be ruled out. For the absorbed APEC fit, sources 3 and 5 have unphysical kT values ($>9$ and $>18$~keV) and source 6 gives an unsatisfactory fit. For the absorbed blackbody or power law fits for the sources for which coronal emission can be ruled out, source 5 has an $N_{\rm H}$ value too low to be associated with the remnant, and this is also true for the blackbody fit for source 6. Source 6 is well fitted with a power law at the distance of the remnant. However, the power law index of $5.9\pm0.9$ is very steep, ruling source 6 out as a neutron star candidate. 

\section{Timing}\label{timing}
Because the pn observation was done in full window mode, the timing resolution
of 70~ms limits pulsar searches to frequencies below approximately 7~Hz, 
which would exclude all millisecond pulsars and many ordinary pulsars, but
may identify slowly rotating magnetars known to have periods $P\sim2-12$~s. The 
maximum frequency is applicable only to pulsars with an approximately 
sinusoidal lightcurve (i.e. a single peak per rotation). Pulsars with 
lightcurves which have additional peaks or other features would likely not 
be found unless the frequency was much less than 3.6~Hz. A negative result in
a timing search thus does not indicate with any certainty that there is no
pulsar present.

While producing the events file, the SAS software showed a possible problem with the timing for quadrant 3, involving jumps in the course timing parameter. This means that the timing for sources 3 and 4 may not have been reliable, and there is no way to fix the timing with the data from this observation. Timing data from the other numbered sources would not have been affected by this.

Blind (i.e. the frequency is not known beforehand) timing searches were 
performed on the point sources of interest, including sources 3 and 4. For these searches, an FFT is 
first calculated, and then frequencies surrounding any significant peaks 
are searched using Rayleigh \citep{lea83} (also known as $Z_1^2$), $Z_2^2$, $Z_3^2$ \citep{buc83},
and epoch folding tests. Only searches for frequencies of less than 1.8~Hz
would have a sufficient number of bins per cycle (8) to reliably perform 
most of the tests. 

\textbf{None of the point sources surrounding SNR \g107\ showed evidence of 
detectable pulsations. }

\section{Discussion}\label{disc}
Because of their proximity to the centre of the radio shell, and because they can
be fitted with an $N_{\rm H}$ value which conforms to the value for the
SNR, sources 1 and 2 are of particular interest for the analysis presented in this paper. 
All other sources investigated in this study are also ruled out as neutron star candidates associated with this remnant, because of their proximity to the shell, their fitted value of $N_{\rm H}$, and/or the lack of conformity between the individual spectra and both a power law and a blackbody. 

Source 1 is a soft thermal source which is best fitted with an 
absorbed blackbody with a 
temperature of $(9.5\pm0.5)\times 10^5$~K, when 
$N_{\rm H}$ is frozen at $0.6\times 10^{22}$~cm$^{-2}$. Its corresponding
luminosity, assuming a distance of 1.1~kpc, is $(3.0^{+1.0}_{-0.7})\times 10^{32}$~erg s$^{-1}$.
An absorbed power 
law fit to the spectrum of source 1 yields an unacceptably high value of $\chi^2_\nu$, 
indicating that its spectrum is dominated by a thermal component. 
These values are qualitatively comparable with the spectral parameters 
and luminosity of the Vela pulsar \citep{pav01}, and the age of 3--6~kyr is close to Vela's 
characteristic age of 11~kyr, indicating that source 1 may be a Vela-like 
pulsar. Alternatively, this source could be a CCO candidate, since the X-ray spectra of CCOs are dominated by a thermal blackbody spectrum and have comparable luminosities \citep[see e.g.][]{pav04}. The spectral analysis did not rule out the possibility that the X-ray emission from source 1 arises from the corona of a late-type star, however, and since it is unlikely to find such a bright star by chance, this likely rules out Source 1 as a neutron star or CCO.

The fitted spectral parameters and proximity to the centre of the SNR radio shell of source 2 are similar to those of source 1, and it thus has similar merit as a neutron star candidate. Similarly, coronal emission is not ruled out as producing the X-ray emission from source 2, and since it does have a close optical and infrared counterpart, it is also likely ruled out as a neutron star or CCO. 

Source 1 appears in the \ro\ All-Sky Survey catalogue, but source 2 does not. Since source 2 has a flux similar to that of source 1, according to the present X-ray data, it is possible that it fell just short of inclusion in the RASS. Sources 1 and 2 are included in the \xmm\ serendipitous source catalogue \citep{wat09} as 2XMMi J225200.3+574259 and J225236.3+574422, respectively. 



A deeper observation with higher angular resolution, such as with \cha, would better allow closely-spaced X-ray sources to be distinguished, and would permit further spectroscopic study of the individual sources, particularly those that are too faint to have been studied in this paper. With more spectroscopic counts, multicomponent models could be used for the sources that do not fit well to simpler models, and source identification would be facilitated. In addition, with smaller X-ray error circles, association of X-ray sources with possible optical counterparts, such as that of source 1 with the nearby bright optical source, could be determined less ambiguously.

Timing observations for the as yet unidentified sources are also needed, and due to their proximity to each other, it would be possible to make a single \xmm\  observation in small window mode which would include both sources and provide the timing resolution necessary for a pulsar search.

For X-ray detection of the diffuse emission from the SNR, a deeper \xmm\ observation near the center of the radio shell and including many of the point sources (indicated with magenta circles on Figure~\ref{xrayradioimageps}) for which spectral analysis was not possible with the available data, would be best. It is possible that, without the reduction in integration time caused by the proton flaring, diffuse X-ray emission from the SNR would have been detected, so another \xmm\ observation of the same duration or longer would be beneficial for a greater understanding of this low-surface brightness, highly polarized SNR.

\section{Acknowledgments}

{\it XMM-Newton} is an ESA science mission with instruments and
contributions directly funded by ESA Member States and NASA. This
research was supported by Natural Sciences and Engineering Research Council of Canada (NSERC). SSH acknowledges support by NSERC, the Canadian Space Agency, Canada Foundation for Innovation, 
and the Canadian Institute for Theoretical Astrophysics.
This research made use of NASA's Astrophysics Data System
and the High Energy Astrophysics Science Archive Research Center (HEASARC). 
We would like to thank the anonymous referee for comments that helped to improve the paper.

\end{document}